\documentclass[runningheads]{styles/svmult}

\usepackage{mathptmx}       
\usepackage{helvet}         
\usepackage{courier}        
\usepackage{type1cm}        
\usepackage{makeidx}         
\usepackage{graphicx}        
\usepackage{multicol}        
\usepackage[bottom]{footmisc}
\usepackage{booktabs} 
\usepackage{subcaption}

\makeindex             

\begin{document}

\title*{Real-Time Agile Software Management for Edge and Fog Computing Based Smart City Infrastructure}
\titlerunning{Agile S/W Management for Edge \& Fog Computing Based Smart City Infrastructure}

\author{Debasish Jana\thanks{Corresponding Author} \and
Pinakpani Pal \and
Pawan Kumar \thanks{The opinions presented in this paper are entirely those of the authors and do not necessarily represent the official positions of their respective institutions.}}


\institute{
Debasish Jana\textsuperscript{*} \at Department of Computer Science and Engineering, Heritage Institute of Technology, Kolkata, India \email{debasish.jana@heritageit.edu}
\and
Pinakpani Pal \at Electronics \& Communication Sciences Unit (ECSU), Indian Statistical Institute (ISI), Kolkata, India \email{pinak@isical.ac.in} 
\and
Pawan Kumar \at Town and Country Planning Organisation, New Delhi, India \email{pawan612@gmail.com} 
}



%
%
%
\maketitle
\vspace{-25mm}
\abstract{The evolution of smart cities demands scalable, secure, and energy-efficient architectures for real-time data processing. With the number of IoT devices expected to exceed 40 billion by 2030, traditional cloud-based systems are increasingly constrained by bandwidth, latency, and energy limitations. This paper leverages the ROOF (Real-time Onsite Operations Facilitation) framework with decentralized computing at intermediary fog and peripheral edge network layers to reduce latency by processing data near its point of origin. ROOF features fog caching to avoid redundancy, ultra-low-power wireless transmission for energy savings, and AI-driven resource allocation for efficiency. Security is enhanced through TLS encryption, blockchain-based authentication, and edge-level access control. Case studies from Bhubaneswar, Barcelona and Copenhagen validate the use of ROOF in traffic systems and environmental monitoring. The paper concludes by outlining key challenges and prospects of AI-driven analytics in smart urban infrastructure.}

\section{Introduction}
\label{sec:intro}
\vspace{-0.5em} 
Smart cities require advanced software systems to manage complex infrastructures. Integrating Internet of Things (IoT) devices, automated transportation, smart grids, and real-time analytics enhances efficiency and sustainability~\cite{rehan2023internet}. However, these systems pose challenges in software management, security, and scalability. Traditional cloud architectures suffer from high latency and network congestion, limiting mission-critical applications. intermediary and peripheral computing~\cite{syed2025artificial} mitigate these issues by processing data closer to the source, improving responsiveness and reducing bandwidth constraints. According to IoT Analytics ~\cite{IoTAnalytics2024}, the number of connected IoT devices will reach approximately 40 billion by 2030. Managing such volumes requires robust architectures~\cite{gour2023perspective} ~\cite{shahzad2024fair}. ROOF (Real-Time Onsite Operations Facilitation) ~\cite{ahammad2021software} provides a decentralized, scalable, and secure approach to resource allocation and real-time decision making. This paper outlines its layered architecture, case studies, security mechanisms, and role in advancing sustainable and smart environments.

\noindent\textbf{Contributions.} 
This paper presents ROOF architecture integrating intermediary fog and peripheral edge computing for real-time data processing. It optimizes resource allocation through sensor aggregation and caching while reducing network congestion. The applicability of ROOF is validated through smart traffic control and environmental monitoring case studies. A security framework integrating encryption, authentication, and blockchain-based access control enhances integrity and privacy. The paper explores the scalability and connectivity of ROOF in smart cities, highlighting future advancements.

\noindent\textbf{Paper Organization.} 
Section~\ref{sec:smarturbanplan} delves into the fusion of modern information \& communication technologies (ICT) in urban infrastructure, addressing service improvements and scalability challenges. Section~\ref{sec:feasibleimpl} presents ROOF as a scalable smart city framework, leveraging fog and cloud-based storage, analytics, and microservices for efficient service integration. Section~\ref{sec:EssenceAgile} explores the integration of Agile with the Essence Kernel for better project tracking and collaboration.  Section~\ref{sec:casestudy} discusses smart city initiatives, focusing on e-Governance and global trends. Finally, Section~\ref{sec:concl} concludes with future directions in scalability, security, and open standards.
\vspace{-0.5em} 
\section{Integration of Modern Technologies in Smart Urban Infrastructure}
\label{sec:smarturbanplan}
\vspace{-0.5em} 
Smart cities~\cite{okonta2024smart} leverage the convergence of Internet of Things, cloud infrastructures, artificial intelligence, and data-driven analytics to optimize sustainable urban operations. Cities such as Amsterdam, Singapore, and Bhubaneswar~\cite{niua2023smartcities} have successfully integrated these technologies, yet large-scale adoption faces challenges in scalability, interoperability, and security~\cite{jana2024smartcity1}.

\noindent\textbf{Key Stakeholders in Smart City Development. } Smart city initiatives require collaboration between \textit{local governments}, \textit{technology providers}, \textit{urban planners}, \textit{regulatory bodies}, and \textit{citizens}. Public-private partnerships (PPPs) share financial and technological responsibilities, as seen in India’s Special Purpose Vehicle (SPV) model, where \textit{Urban Local Bodies (ULBs)}, state governments, and private companies work together to balance innovation with public interests.

\noindent\textbf{Key Technologies in Smart City Infrastructure. }Several cutting-edge technologies are integral to the successful operation of smart cities:

\begin{itemize}
\vspace{-2mm}
\item \textbf{IOT: } \textit{IoT} devices enable real-time urban monitoring, improving traffic flow, energy management, and public safety~\cite{pawan2021iot}. For instance, Barcelona’s smart waste bins optimize collection routes, reducing costs, while adaptive traffic systems minimize congestion and emissions.

\item \textbf{Big Data Analytics: }By analyzing data from IoT sensors, public services, and social media, cities enhance resource allocation and infrastructure planning. Singapore leverages real-time analytics to prevent flooding and manage water distribution efficiently.

\item \textbf{Cloud, Fog, and Edge Computing: }These technologies provide scalable infrastructure for smart city data management. While cloud computing supports vast storage and analytics, edge and fog computing bring computation closer to the data source to reduce response times. Los Angeles, for example, employs cloud solutions for open data platforms, offering real-time traffic and crime updates.

\item \textbf{Artificial Intelligence(AI) powered insights using machine learning (ML): } AI-powered analytics improve decision making in smart cities, optimizing public transport schedules, energy consumption, and environmental monitoring. Tokyo’s AI-powered transit system enhances efficiency by predicting train arrival times, reducing delays.

\item \textbf{Mobile Computing: }Mobile applications offer seamless access to city services, improving user engagement. However, security and bandwidth limitations remain key concerns in ensuring reliable service delivery.

\end{itemize}

\noindent\textbf{Applications in Urban Mobility. }Modern technologies optimize urban transportation by integrating real-time data from vehicles, public transit, and traffic systems:
\begin{itemize}
\vspace{-2mm}
\item\textbf{Real-time Information Dissemination: }IoT-enabled traffic updates improve commuter experience, as seen in Singapore’s real-time bus tracking system.

\item\textbf{Route Optimization: }IoT sensors monitor traffic to adjust transit schedules dynamically, enhancing efficiency in high-demand areas.

\item\textbf{Connected Vehicles \& Smart Traffic Systems: }Communication between vehicles (\textit{V2V}) and with roadside infrastructure (\textit{V2I}) enhances traffic efficiency and promotes safer driving conditions. Amsterdam’s adaptive traffic signals leverage V2I to minimize congestion.

\end{itemize}

\noindent\textbf{Challenges in Integrated Software Management. }The key challenges include:

\begin{itemize}
\vspace{-2mm}
\item\textbf{Scalability: }Growing IoT adoption increases complexity. Microservice architecture (MSA) improves scalability by enabling modular, independent services~\cite{ray2020extending}.

\item\textbf{Interoperability: }Proprietary systems hinder seamless integration. Open standards and APIs facilitate effective data transmission.

\item\textbf{System Security and User Privacy: }Cybersecurity risks necessitate strong data encryption, secure access mechanisms, and threat detection systems to safeguard critical infrastructure.

\item\textbf{Real-Time Data Handling: }Distributed data handling needs efficient synchronization in timely decisions, especially in transportation and emergency services.
\end{itemize}

\noindent By addressing these challenges, smart cities can leverage IoT, AI, and cloud technologies to design cities that promote sustainability, operational efficiency, and enhanced livability.

\section{ROOF Architecture: A Scalable Smart City Framework}
\label{sec:feasibleimpl}

\begin{figure}[t!]
    \centering
    \begin{subfigure}[t]{0.32\textwidth}
        \centering
        \includegraphics[width=\textwidth]{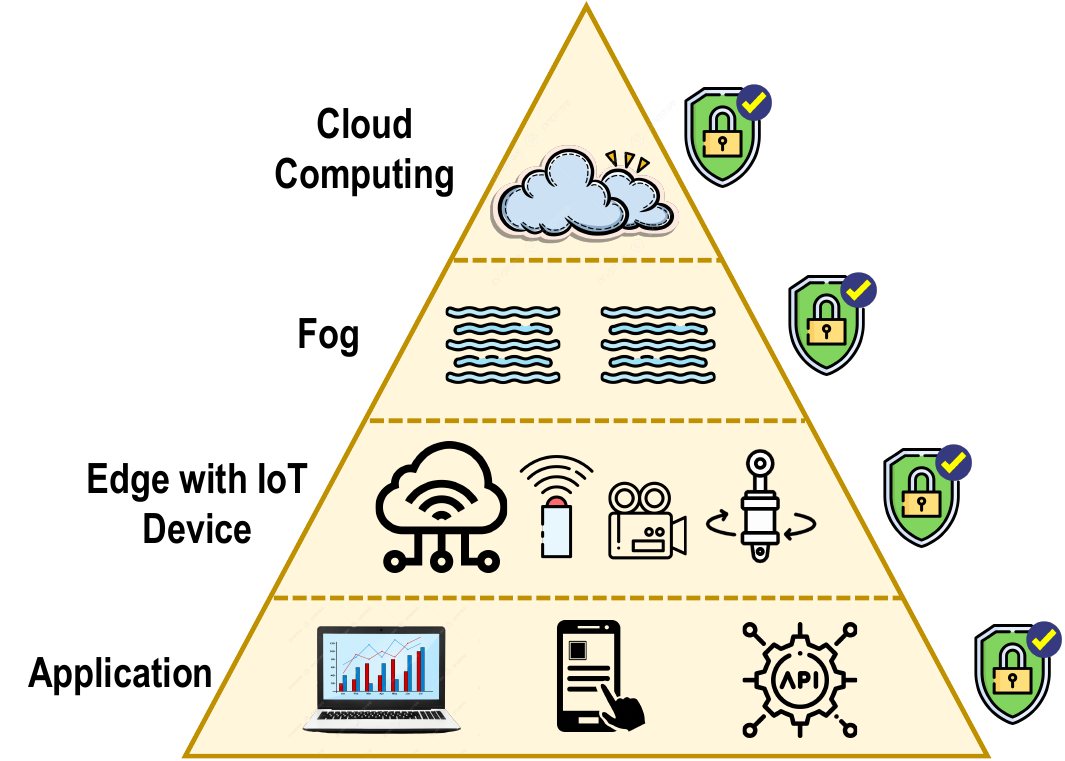}
        \caption{ROOF architecture}
        \label{subfig:archlayers}
    \end{subfigure}%
    \hspace{2mm}
    \begin{subfigure}[t]{0.63\textwidth}
        \centering
        \includegraphics[width=\textwidth]{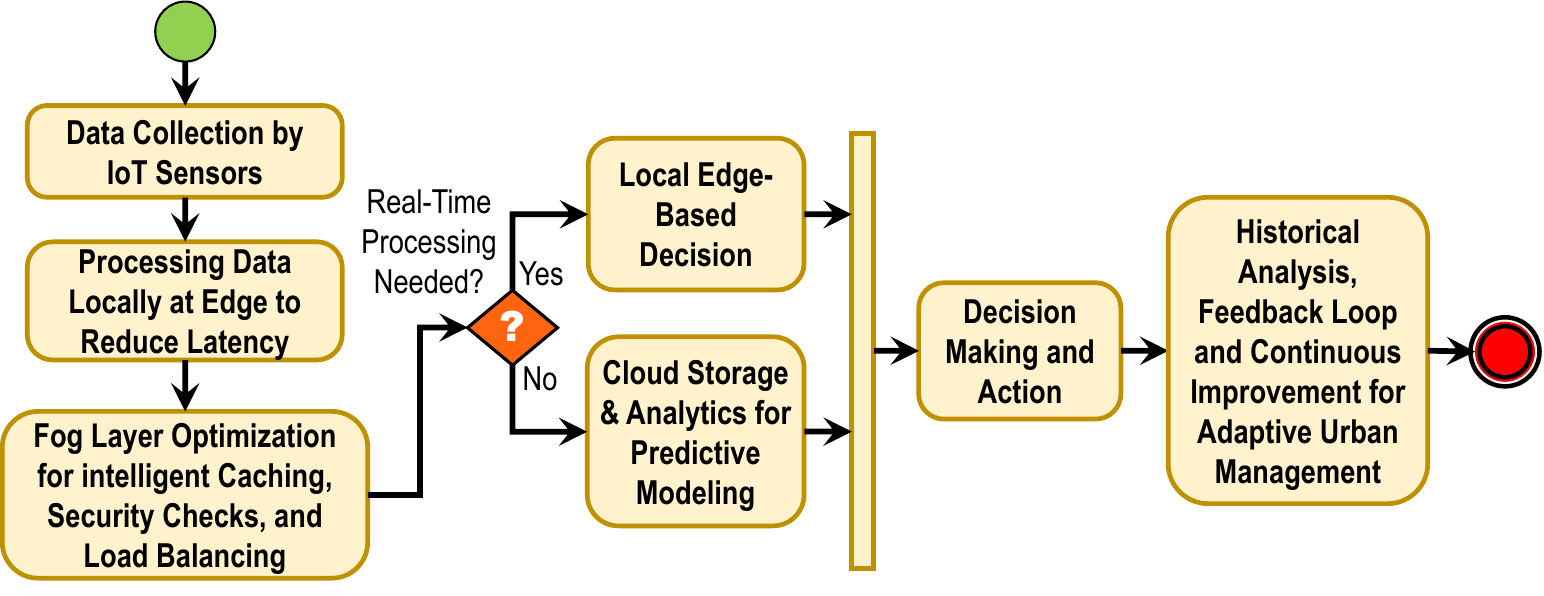}
        \caption{Workflow in ROOF framework}
        \label{subfig:flowchart01-horz}
    \end{subfigure}
    \caption{Overview of the ROOF architectural framework and workflow}
    \vspace{-2mm}
\end{figure}  

Existing smart city architectures struggle with interoperability, security, and real-time processing. The integration of diverse IoT sensors and multiple stakeholders requires a robust framework for data sharing, secure communication, and adaptive resource allocation. ROOF ~\cite{ahammad2021software} addresses these challenges with a hierarchical, multi-layered microservices-based architecture (MSA) that enhances flexibility and scalability. Each microservice can focus on a specific function—such as traffic monitoring, energy management, or environmental data collection—without affecting the entire system. The ROOF architecture (Fig. \ref{subfig:archlayers}) consists of the following layers:
\begin{itemize}
\vspace{-2mm}
\item \textbf{Edge with IoT Device Layer :} IoT sensors, cameras, and actuators collect real-time urban data and are able to transmit essential information to fog nodes.

\item \textbf{Fog Layer : }
Intermediate fog nodes perform local analytics, caching and security validation, minimizing latency and optimizing cloud communication. Technologies like Apache Kafka, Spark enable real-time streaming and processing.

\item \textbf{Cloud Layer : }
Centralized storage and analytics handle long-term data processing, using microservices for smart city functions, and integrating platforms such as Microsoft Azure and AWS for scalable city services.

\item \textbf{Application Layer : }
Provides end-user functionality, like dashboards, citizen mobile apps, and developer APIs for urban monitoring and management.

\item \textbf{Security Layer : }
Implements encryption, authentication, and blockchain-based access control at every layer to ensure data privacy and regulatory compliance.

\setlength{\tabcolsep}{2pt} 
\begin{table}[t]
\centering
\caption{Performance Comparison: ROOF vs. Cloud}
\label{tab:cloudvsroof}
\vspace{-2mm}
\begin{tabular}
{@{}p{2.6cm}p{4.5cm}p{4cm}@{}} 
\toprule
\textbf{Metric} & \textbf{Traditional Cloud} & \textbf{ROOF Framework}\\ 
\midrule
Latency & High due to centralized processing & Low with edge computing \\
Energy Efficiency & High power use from redundancy & Optimized via fog computing \\
Scalability & Limited by network congestion & Modular \& adaptable for IoT \\
Security \& Privacy & Centralized, prone to breaches & Decentralized, blockchain-based\\
Real-Time & Slower due to remote centers & Faster with local processing \\
\bottomrule
\end{tabular}
\vspace{-6mm}
\end{table}
\end{itemize}
The ROOF methodology (Fig.~\ref{subfig:flowchart01-horz}) illustrates the flow from data collection to real-time decision making. IoT sensors gather data from traffic, environmental, and monitoring systems. Local edge processing reduces latency before forwarding data to fog nodes for security validation, caching, and load balancing. Non-critical data is sent to cloud storage for historical analysis and predictive modeling. Decentralized analytics optimize resource allocation, prioritizing critical data for minimal delay. Table \ref{tab:cloudvsroof} lists the performance benefits of the ROOF framework from the traditional cloud. The technology stack comprises the following components:

\begin{itemize}
\vspace{-2mm}
\item \textbf{Programming Languages: } Java, Python (backend), JavaScript (frontend).
\item \textbf{Frameworks: } Spring Boot (microservices), Node.js (real-time UI).
\item \textbf{Data Management: } MongoDB, Cassandra (DB), Apache Kafka (streaming).
\item \textbf{Cloud Platforms: } AWS, Azure (scalable deployment).
\item \textbf{Security Protocols: } TLS/SSL (secure communication), OAuth 2.0 (authentication), firewalls (network security).
\vspace{-2mm}
\end{itemize}
The key features of ROOF~\cite{krishna2023white} include multi-sensor virtualization for adaptive resource allocation, fog-based caching to reduce redundancy, and real-time prioritization for responsiveness. The architecture supports applications such as traffic monitoring, air quality analysis, and energy management, making it scalable.

\section{Essence Kernel and Agile Process in Software Development}
\label{sec:EssenceAgile}
\begin{figure}[t!]
    \centering
    \begin{subfigure}[t]{0.25\textwidth}
        \centering
        \includegraphics[width=\textwidth]{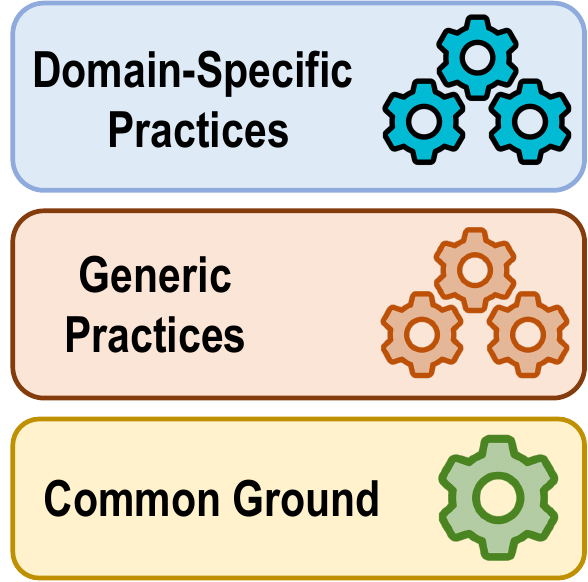}
        \caption{Essence Layers}
        \label{subfig:EssenceArch}
    \end{subfigure}    
    \hspace{0.1mm}
    \begin{subfigure}[t]{0.42\textwidth}
        \centering
        \includegraphics[width=\textwidth]{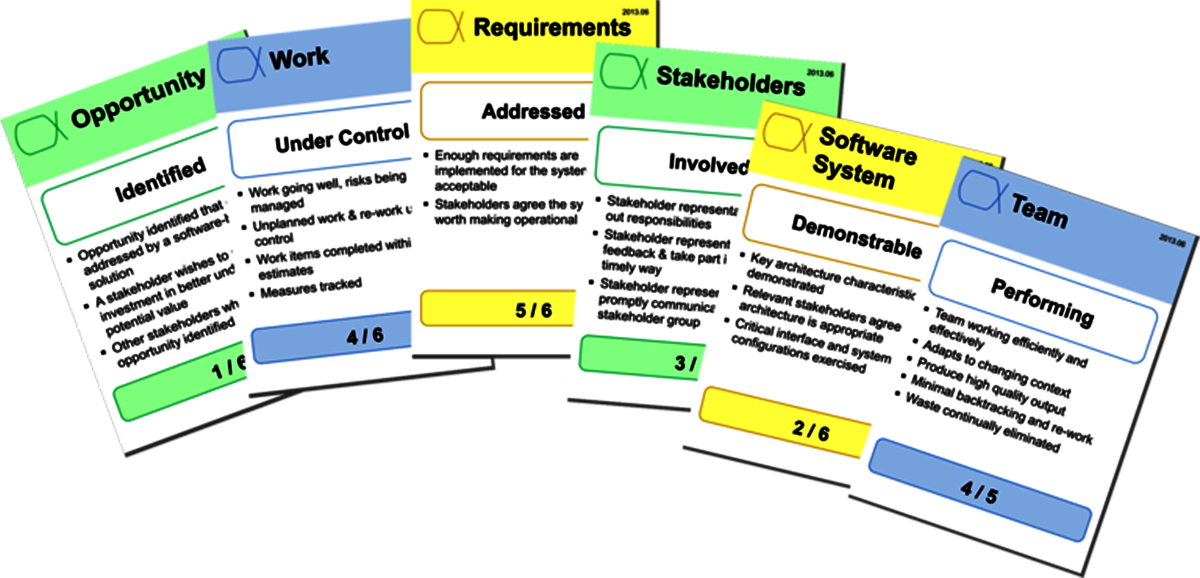}
        \caption{Alpha Cards}
        \label{subfig:alphacardsColor}
    \end{subfigure}
    \hspace{0.1mm}
    \begin{subfigure}[t]{0.3\textwidth}
        \centering
        \includegraphics[width=\textwidth]{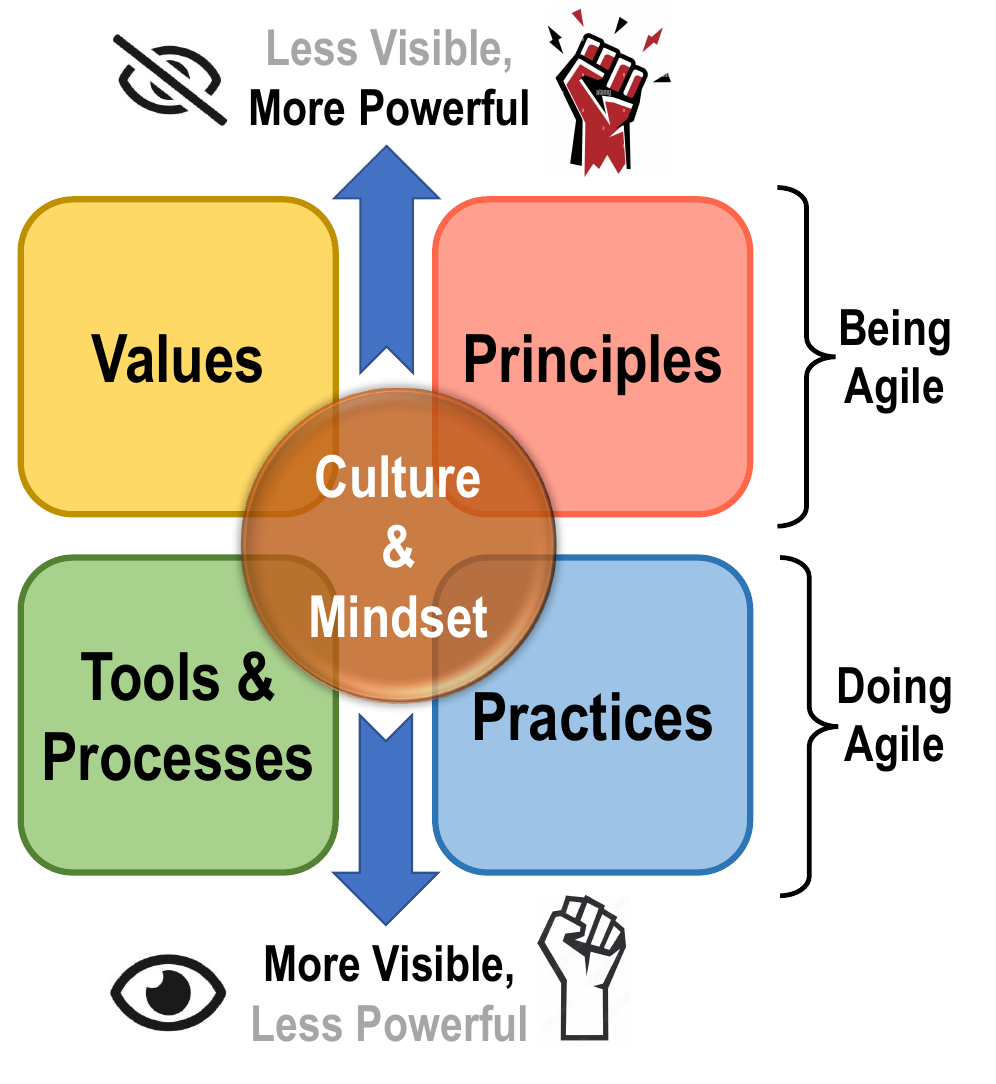}
        \caption{Agile Process}
        \label{subfig:AgileProcessArchColor}
    \end{subfigure}
    \caption{Agile practices and Essence improvisation}
    \vspace{-6mm}
\end{figure}

\noindent \textbf{Essence Kernel}, by Software Engineering Method and Theory (SEMAT) initiative and standardized by the Object Management Group (OMG), offers a method-agnostic adaptable framework to track progress and identify issues~\cite{jacobson2013essence}. Centered on \textit{Alphas}—core elements like stakeholders and requirements—each progressing through defined states from the inception to deployment. The Essence framework (Fig. \ref{subfig:EssenceArch}) structures generic and domain-specific practices atop the kernel.

\noindent\textbf{Alpha Cards} (Fig.~\ref{subfig:alphacardsColor}) serve as visual tools to track the progress of the project. The \textit{Opportunity Alpha} ensures clarity on business value to be clearly understood by the stakeholders, while the \textit{Software System Alpha} tracks the evolution of the software, from concept to deployment. Ericsson used Alpha Cards to standardize project oversight across distributed teams, improving coordination and efficiency.

\noindent\textbf{Agile Methodology} (Fig.~\ref{subfig:AgileProcessArchColor}) enables iterative development, allowing teams to deliver software in increments through short iterations called sprints~\cite{chathuranga2023practices}. Agile fosters collaboration, ensuring quick adaptation to changing customer needs and technological progress. Continuous integration, automated testing, and DevOps enhance Agile workflows, simplifying software deployment. Companies like Spotify use Agile’s small, autonomous squads to develop and refine product features efficiently.

\noindent\textbf{Integrating Essence Kernel with Agile Methodology} balances flexibility with structured progress tracking. Agile excels in adaptability, but lacks a formal mechanism for assessing overall project health. The Essence Kernel provides this, ensuring alignment across large-scale initiatives. Munich Re, a global reinsurance firm, combined Agile with Essence Kernel to manage digital transformation~\cite{mcdonough2014munich}, leveraging Agile for rapid iteration and Essence for structured oversight through Alpha Cards.

\noindent\textbf{Scaling Agile in Large Enterprises} requires coordination between multiple teams. The Essence Kernel improves scalability by offering a unified project view, minimizing bottlenecks, and ensuring alignment on key deliverables. ING Netherlands successfully implemented Agile with the Essence Kernel for its digital banking needs, balancing frequent updates with security, compliance, and user experience.

\noindent\textbf{Debunking Agile Myths} (Table \ref{tab:agilemyths}) is essential for effective adoption. Agile does not inherently reduce development time or eliminate documentation~\cite{jana2020essence}. Instead, it requires careful planning, structured design, and proper documentation to maintain software quality. Facebook and Google emphasize documentation within Agile workflows to ensure scalability, feature consistency, and regulatory compliance.

\setlength{\tabcolsep}{2pt} 
\begin{table}[t]
\centering
\caption{Myths and Facts about Agile Adoption}
\label{tab:agilemyths}
\vspace{-2mm}
\begin{tabular}{@{}p{4.8cm}p{7.2cm}@{}} 
\toprule
\textbf{Myth} & \textbf{Reality} \\ 
\midrule
Agile reduces development time & Speed depends on team expertise \& project complexity \\ 
Agile means no documentation & Documentation is crucial for maintaining quality \\ 
Agile eliminates design phases & Agile needs flexible, scalable designs for iterations \\ 
Agile is always faster & Allows quicker iterations, but the total timeline varies\\ 
Self-organizing teams reduce managers & Managers support and manage team dependencies\\ 
\bottomrule
\end{tabular}
\vspace{-6mm}
\end{table}

\noindent\textbf{Essence and Agile in Smart Cities} create a scalable framework for managing complex urban systems. Agile allows quick adaptation to evolving requirements, while Alpha Cards provide structured monitoring of security, scalability, and performance. Companies like Ericsson, Munich Re, and ING demonstrate the effectiveness of this integration. As smart cities evolve, combining Agile's flexibility with Essence's structured governance will be critical to meet the demands of urban infrastructure.

\section{Real-World Examples and Case Studies}
\label{sec:casestudy}
\vspace{-4mm}
This section explores how IoT-driven smart city initiatives enhance urban living, focusing on Bhubaneswar’s e-governance advancements~\cite{sahoo2019smart} and global implementations in smart traffic control, smart grids, and environmental monitoring.

\noindent\textbf{Bhubaneswar: A Model Smart City in India. }
\noindent\textit{Bhubaneswar}, Odisha’s capital, has emerged as a leading smart city~\cite{BhubaneswarSmartCity2017}, earning top honors in the Smart City Challenge and awards for its innovative e-governance and sustainable urban solutions. Through the Smart Cities Mission~\cite{niua2023smartcities}, Bhubaneswar leverages smart technologies such as real-time data-driven traffic management~\cite{CRUT2024} to drive economic growth and enhance citizens' quality of life. Key initiatives include:
\begin{itemize}
\vspace{-2mm}
\item \textbf{Smart Traffic Management:} \textit{Automatic Number Plate Recognition (ANPR)} detects violations, \textit{e-challan systems} automates fines, and \textit{Adaptive Traffic Signal Control Systems (ATSCS)} optimizes signals for safety and reduced congestion. 
\item \textbf{Traffic Analytics \& Monitoring:} \textit{Automatic Traffic Counting and Classification (ATCC)} measures vehicle flow, while \textit{Traffic Accident Recording Systems (TARS)} use IoT-enabled distance measurement for accident prevention ~\cite{verma2019modernization}.
\item \textbf{Pedestrian and Commuter Safety:} \textit{Pelican crossings} and \textit{Dynamic Message Signboards (DMS)} enhance pedestrian mobility and real-time traffic updates.
\item \textbf{IoT-Based Environmental Monitoring (EVMS):} Sensor stations track air quality, noise, temperature, and humidity, for efficient data sharing among agencies.
\end{itemize}

\noindent A prototype implementation of ROOF demonstrates edge-based analytics for real-time traffic predictions, adaptive signals, and emergency routing. Key outcomes include (a) edge AI for congestion prediction, (b) fog-based dynamic signal control, and (c) distributed decision-making for emergency vehicle priority routing, validating its effectiveness in smart urban mobility and environmental monitoring.

\noindent\textbf{Global Smart City Innovations. } Some of the key initiatives include:
\begin{itemize}
\label{subsec:smartworldwide}
\vspace{-2mm}
\item\textbf{Smart Traffic Control: }\textit{Singapore} and \textit{Barcelona} use IoT sensors for congestion management~\cite{quadar2021smart}, reducing travel times by 21\% and emissions by 30\%. Singapore's system~\cite{niculescu2015smart} also monitors water quality for environmental sustainability. 
\item\textbf{Smart Grids \& Energy Efficiency: }\textit{San Diego}'s smart grids optimize energy use and support EV charging, stabilizing power during peak demand~\cite{sultan2022integration}.
\item\textbf{Environmental Sustainability: }\textit{Copenhagen} uses IoT~\cite{channi2022role} for smarter infrastructure—managing lighting, waste, and environmental metrics in real time. \textit{Amager Bakke} converts over 560,000 tons of waste into energy yearly, while citywide sensors help monitor air and water quality for prompt sustainable action~\cite{bisinella2022environmental}.
\end{itemize}

\section{Conclusion and Future Work}
\label{sec:concl}
\vspace{-2mm}
ROOF offers a decentralized and scalable architecture for processing smart city data in real time, leveraging intermediary fog and peripheral edge computing, virtualized multi-sensor networks, and streamlined security protocols. Built on a microservices framework, it promotes modularity and fault tolerance. With cloud-edge integration, it supports efficient urban data handling by minimizing latency and improving network performance. Future enhancements will integrate machine learning for predictive analytics and extend ROOF to Smart Villages for low-cost sensing, community-driven data collection, and sustainable energy solutions. Addressing challenges in scalability, interoperability, and data security requires coordinated efforts from government, industry, and citizens. The adoption of open standards, blockchain-based security, and AI-driven analytics will further strengthen predictive urban management, ensuring adaptability and security in an evolving regulatory landscape.
\vspace{-6mm}
%
%
\bibliographystyle{styles/spmpsci}
\bibliography{files/mybib}

\end{document}